\documentclass[preprint,showpacs,preprintnumbers,amsmath,amssymb]{revtex4}
\usepackage{bm}
\usepackage{epsf}

\everymath{\displaystyle}
\begin{document}           
\title{Classical Limit of Relativistic Quantum Mechanical Equations
in the Foldy-Wouthuysen Representation}

\author{A. J. Silenko}\email{silenko@inp.bsu.by}
\affiliation{Research Institute for Nuclear Problems, Belarusian
State University, Minsk, 220030 Belarus}

\begin{abstract}
It is shown that, under the Wentzel-Kramers-Brillouin
approximation conditions, using the Foldy-Wouthuysen
representation allows the problem of finding a classical limit of
relativistic quantum mechanical equations to be reduced to the
replacement of operators in the Hamiltonian and quantum mechanical
equations of motion by the respective classical quantities.
\end{abstract}

\pacs {03.65.Pm, 03.65.Sq, 11.10.Ef} \maketitle


The Foldy-Wouthuysen (FW) representation \cite{FW} possesses
unique features that make it special in quantum mechanics. Even
for relativistic particles in the external field, the operators in
this representation are completely analogous to the respective
operators of nonrelativistic quantum mechanics. In particular, the
operators of position \cite{NW} and momentum are $\bm r$ and $\bm
p=-i\hbar\nabla$, while that of polarization for half-spin
particles is expressed by the Dirac matrix $\bm \Pi$. In other
representations these operators are given by much more awkward
formulas (see \cite{FW,JMP}). The simple and well-defined form of
operators corresponding to classical observables is a major
advantage of FW representation. Note that in this representation
the Hamiltonian and all operators are diagonal in two spinors
(block-diagonal). The usage of the FW representation eliminates
the chance that ambiguities will occur while solving the problem
of finding a classical limit of relativistic quantum mechanics
\cite{FW,CMcK}.

In the nonrelativistic case, a transition to the quasiclassical
approximation is relatively easily done using
Wentzel-Kramers-Brillouin method (WKB). It can be applied when a
de Broglie wavelength is smaller than the characteristic size of
the inhomogeneity region of the external field $l$:
\begin{equation}
\lambda\ll l. \label{eqdeBro}\end{equation} For a one-dimensional
problem (motion along $x$ axis only) from (\ref{eqdeBro}) it
follows that the following inequality holds:
\begin{equation}
\left|\frac{d\lambda}{dx}\right|\ll1. \label{eqderdeBro}
\end{equation}

It is also convenient to use the WKB method in the analysis of
relativistic quantum mechanical equations. However, in this case
it needs to be modified. Like in nonrelativistic quantum
mechanics, a classical limit is reached in the zero-order WKB
approximation in $\hbar$. If the FW representation is used, a
transition to the quasiclassical approximation is done in the same
way as in nonrelativistic quantum mechanics. In the relativistic
case, in order to be able to use the WKB method, conditions
(\ref{eqdeBro}) and (\ref{eqderdeBro}) must also be satisfied.

To simplify the analysis, let us consider the case of
one-dimensional motion. If no spin effects are taken into account,
then the equation for relativistic Hamiltonian in the FW
representation can be written as follows:
\begin{eqnarray}
i\hbar\frac{\partial \Psi}{\partial t}={\cal H}\Psi, ~~~ {\cal
H}=\sqrt{m^2c^4+c^2\bm p^2+{\cal V}(x,\bm p)}+U(x). \label{fVKBn}
\end{eqnarray}
This form has, in particular, an equation for scalar particles in
the electromagnetic field (see \cite{TMP2008}). Since the
operators $p_y,p_z$ commute with the Hamiltonian and have definite
values, the operator ${\cal H}$ can be represented in the form
\begin{eqnarray}
i\hbar\frac{\partial \Psi}{\partial t}={\cal H}\Psi, ~~~ {\cal
H}=\sqrt{m^2c^4+c^2p_x^2+V(x,p_x)}+U(x). \label{fVKB}
\end{eqnarray}

For stationary states, a usual form for the wave function can be
used (see \cite{Dav,LL}):
\begin{eqnarray}
\Psi=\exp{\left(-\frac{i}{\hbar}Et\right)}\Phi(x), ~~~
\Phi(x)=\exp{\left(\frac{i}{\hbar}\mathfrak{S}\right)}, \label{pdnvBKB}
\end{eqnarray}
where $E$ is the total energy of a particle. The function
$\mathfrak{S}$ can formally be expanded into a series in powers of
the Planck constant:
\begin{eqnarray}
\mathfrak{S}=\mathfrak{S}_0+\frac{\hbar}{i}\mathfrak{S}_1+\left(
\frac{\hbar}{i}\right)^2\mathfrak{S}_2+\ldots. \label{developVKB}
\end{eqnarray}

When substituting the wave function into the initial equation, we
limit ourselves by the terms of zero-order approximation. In the
latter approximation, the commutators of $x$ and $p_x$ operators
proportional to $\hbar$ can be neglected. Since
\begin{eqnarray}
\bm p^2\Psi=({\mathfrak{S}'}^2-i\hbar\mathfrak{S}'')\Psi,
\label{pVKB}
\end{eqnarray}
ignoring the quantities of the first and higher orders in $\hbar$,
we find
$$\sqrt{m^2c^4+c^2p_x^2+V(x,p_x)}\Psi=
\sqrt{m^2c^4+c^2{\mathfrak{S}'}^2 +V(x,\mathfrak{S}')}\Psi. $$

Thus, the terms of zeroth order in the Planck constant satisfy the
following equation:
\begin{eqnarray}
E=\sqrt{m^2c^4+c^2{\mathfrak{S}'}^2+V(x,\mathfrak{S}')}+U(x),
\label{nlevVKB}
\end{eqnarray}
which defines an implicit function $\mathfrak{S}'$ of $x$. It is
clear that the quantity $\mathfrak{S}'$ is a classical generalized
momentum of a particle ${\cal P}(x)$. Therefore,
\begin{eqnarray}
\mathfrak{S}=\int{{\cal P}(x)dx}. \label{nintVKB}
\end{eqnarray}

Thus, $\mathfrak{S}$ is a time-independent part of the action,
while, according to the initial equation (\ref{fVKB}), the total
action of a particle is found to be
\begin{eqnarray}
{\cal S}=-Et+\mathfrak{S}=-Et+\int{{\cal P}(x)dx}. \label{efinVKB}
\end{eqnarray}

Formula (\ref{efinVKB}) is completely consistent with the
classical theory and coincides with the analogous one deduced for
the WKB approximation in nonrelativistic quantum mechanics
\cite{Dav,LL}. Thus, while using the FW representation in
relativistic quantum mechanics, a transition to the classical
limit corresponds to the zero-order WKB approximation in $\hbar$.
As follows from (\ref{fVKB}),(\ref{nlevVKB})--(\ref{efinVKB}),
this can be done by replacing the operators in the Hamiltonian by
the respective classical quantities.

It is easy to show that such a replacement can be carried out in
equations of motion as well. Any quantum mechanical Hamiltonian is
an operator function of generalized momenta $p_i$ and
corresponding coordinates $x^i$. By neglecting the terms
proportional to $\hbar$, we may not take into account the
noncommutativity of operators of dynamical variables and write a
total time derivative of the Hamiltonian in the form
$$\frac{d{\cal
H}}{dt}=\frac{\partial{\cal H}}{\partial t}+\frac{\partial{\cal
H}}{\partial p_i}\frac{dp_i}{dt}+\frac{\partial{\cal H}}{\partial
x^i}\frac{dx^i}{dt}.$$ Since $$\frac{d{\cal
H}}{dt}=\frac{\partial{\cal H}}{\partial t},$$ then, in zero-order
approximation in $\hbar$, the operator equations of motion can be
represented in a form similar to classical Hamilton equations:
\begin{eqnarray}
\frac{dx^i}{dt}=\frac{\partial{\cal H}}{\partial p_i}, ~~~
\frac{dp_i}{dt}=-\frac{\partial{\cal H}}{\partial x^i}.
\label{eqmtVKB}
\end{eqnarray}
The possibility of replacing the operators in the Hamiltonian by
respective classical quantities according to (\ref{eqmtVKB}) leads
to that of the same replacement in the operator equations of
motion.

There are some peculiarities in applying the WKB method in
gravitation theory \cite{wong3}. However, when using the
Hamiltonian approach (see \cite{PRD,PRD2,OST,OSTRONG}), the
problem of transiting to the classical limit is simplified and
reduced to that considered above. The general form of the
classical Hamiltonian of a spinless particle in arbitrary
electromagnetic and gravitation fields is determined by Eq. (2.5)
given in \cite{Cogn}. It was shown in \cite{OSTRONG} that,
according to the results obtained in \cite{PK} and in order to
describe particles with spin, it should be complemented by the
term $\bm s\cdot\bm\Omega$ proportional to the angular velocity of
spin rotation $\bm\Omega$.

The same term is added to the spinless part of the Hamiltonian
also to describe spin effects in electromagnetic and weak
interactions. Since the Hamiltonian is given in the FW
representation, only the upper spinor can be used. In this case
the operator $\bm s$ is expressed through the spin matrices for
particles with respective spin. For particles with spin $s>1/2$,
the operator ${\cal H}$ may include the products of spin matrices.
After carrying out a transition to the classical limit described
above, the Hamiltonian of particles with spin contains generalized
momenta corresponding coordinates and spin matrices (including
their products). In this case, to find spin dynamics, it is very
convenient to use the method based on the equation for the matrix
Hamiltonian, often called the method of spin amplitudes (see
\cite{PRC} and references therein). A transition to the classical
limit is reduced to averaging the spin matrices and their products
over amplitude spin functions. Such averaging leads to the
introduction of polarization vector $\bm P$ and tensor $P_{ij}$,
given by the equations (see \cite{PRC,MShY})
\begin{eqnarray}
P_i =\frac{<s_i>}{S}, ~~~ P_{ij} = \frac{3 <s_is_j +
s_js_i>-2S(S+1)\delta_{ij}}{2S(2S - 1)}, ~~~ i,j=x,y,z.
\label{Ueq1}\end{eqnarray} Here $s_i$ indicates spin matrices and
$S$ is a spin quantum number.

It should be taken into account that, in relativistic quantum
mechanics, like in nonrelativistic quantum mechanics (see
\cite{LL}), there are some limitations to the use of this WKB
method. The smallness of the discarded term in (\ref{pVKB}), that
contains a higher derivative not always guarantees the smallness
of its contribution to the solution for $\mathfrak{S}(x)$. This
situation can occur when the field extends to distances greater
than the characteristic length $l$, at which it experiences a
noticeable change. A quasiclassical approximation turns then out
to be inapplicable in tracing the behavior of the wave function at
large distances \cite{LL}.

Thus, when the conditions of the WKB approximation are satisfied,
the usage of the FW representation in most cases allows one to
reduce the problem of finding a classical limit of relativistic
quantum mechanical equations to the replacement of operators in
the Hamiltonian and quantum mechanical equations of motion by the
respective classical quantities.

\section* {Acknowledgements}

This work was supported by the Belarusian Republican Foundation
for Fundamental Research, grant No. $\Phi$12D-002.


\begin{thebibliography}{}

\bibitem{FW}
\emph{Foldy L.L., Wouthuysen S.A.} On the Dirac Theory of Spin 1/2
Particles and Its Non-Relativistic Limit // Phys. Rev. 1950. V.
78, No. 1. P. 29-36.

\bibitem{NW}
\emph{Newton T.D., Wigner E.P.} Localized States for Elementary
Systems // Rev. Mod. Phys. 1949. V. 21, Iss. 3. P. 400Ц406.

\bibitem{JMP}
\emph{Silenko A.J.} Foldy-Wouthuysen transformation for
relativistic particles in external fields // J. Math. Phys. 2003.
V. 44, No. 7. P. 2952-2966.

\bibitem{CMcK}
\emph{Costella J.P., McKellar B.H.J.} The Foldy-Wouthuysen
transformation // Am. J. Phys. 1995. V. 63, Iss. 12. P. 1119-1121.

\bibitem{TMP2008}
\emph{Silenko A.J.} Hamilton Operator and the Semiclassical Limit
for Scalar Particles in an Electromagnetic Field
// Theor. Math. Phys. 2008. V. 156, No. 3. P. 1308-1318.

\bibitem{Dav}
\emph{Davydov A.S.} Quantum mechanics (Pergamon, 1965);

\emph{ Levich V.G., Vdovin  Ya.A., Myamlin V.A.}  Theoretical
Physics, Vol.2: Quantum Mechanics. Quantum Statistics and Physical
Kinetics, 2nd ed. (Nauka, Moscow, 1971) [in Russian].

\bibitem{LL}
\emph{Landau L.D. and Lifshits E.M.} Quantum Mechanics:
Non-Relativistic Theory (Pergamon, 1977).

\bibitem{wong3} \emph{Audretsch J.}
Trajectories and spin motion of massive spin-1/2 particles in
gravitational fields // J. Phys. A: Math. and Gen. 1981. V. 14,
No. 2. P. 411-422.

\bibitem{PRD}
\emph{Silenko A.J., Teryaev O.V.} Semiclassical limit for Dirac
particles interacting with a gravitational field // Phys. Rev. D.
2005. V. 71, Iss. 6. P. 064016.

\bibitem{PRD2}
\emph{Silenko A.J., Teryaev O.V.} Equivalence principle and
experimental tests of gravitational spin effects // Phys. Rev. D.
2007. V. 76, Iss. 6. P. 061101.

\bibitem{OST}
\emph{Obukhov Yu.N., Silenko A.J., Teryaev O.V.}  Spin dynamics in
gravitational fields of rotating bodies and the equivalence
principle // Phys. Rev. D. 2009. V. 80, Iss. 6. P. 064044.

\bibitem{OSTRONG}
\emph{Obukhov Yu.N., Silenko A.J., Teryaev O.V.}  Dirac fermions
in strong gravitational fields // Phys. Rev. D.  2011. V. 84, Iss.
2. P. 024025.

\bibitem{Cogn}
\emph{Cognola G., Vanzo L., and Zerbini S.} Relativistic wave
mechanics of spinless particles in a curved space-time // Gen.
Rel. Grav. 1986. V. 18, No. 9. P. 971-982.

\bibitem{PK}
\emph{Pomeranskii A.A., Khriplovich I.B.} Equations of motion of a
spinning relativistic particle in external fields  // J. Exp.
Theor. Phys. 1998. V. 86, Iss. 5. P. 839-849.

\bibitem{PRC}
\emph{Silenko A.J.} Tensor electric polarizability of the deuteron
in storage-ring experiments // Phys. Rev. C. 2007. V. 75, Iss. 1.
P. 014003.

\bibitem{MShY}
\emph{Mane S.R., Shatunov Yu.M. and Yokoya K.} Spin-polarized
charged particle beams in high-energy accelerators // Rep. Prog.
Phys. 2005. V. 68, Iss. 3. P. 1997-2265.

\end{thebibliography}
\end{document}